\documentclass[prc,aps,groupedaddress,preprintnumbers,twocolumn]{revtex4}
\usepackage{graphicx} 

\renewcommand{\vec}[1]{{\mathbf{#1}}}

\begin{document}

\title{Relation between the $0\nu\beta\beta$ and $2\nu\beta\beta$ nuclear matrix elements revisited}

\author{Fedor \v{S}imkovic}
\email[]{fedor.simkovic@fmph.uniba.sk}
\affiliation
        {\it   Laboratory of Theoretical Physics, JINR,
141980 Dubna, Moscow region, Russia
and Department of Nuclear Physics and Biophysics,
Comenius University, Mlynsk\'a dolina F1, SK--842 48
Bratislava, Slovakia}
\author{Rastislav Hod\'{a}k}
\email[] {hodak.rastik@gmail.com}
\affiliation{\it Department of Nuclear Physics and Biophysics,
Comenius University, Mlynsk\'a dolina F1, SK--842 48
Bratislava, Slovakia}
\author{Amand Faessler}
\email[]{amand.faessler@uni-tuebingen.de}
\affiliation{Institute f\"{u}r Theoretische Physik der Universit\"{a}t
T\"{u}bingen, D-72076 T\"{u}bingen, Germany}
\author{Petr Vogel}
\email[]{pxv@caltech.edu}
\affiliation
        {\it Kellogg Radiation Laboratory and Physics Department, 
Caltech, Pasadena, CA 91125, USA}
\date{\today}

\begin{abstract}
We show that the dominant Gamow-Teller part, $M^{0\nu}_{GT}$, 
of the nuclear matrix element governing
the neutrinoless $\beta\beta$ decay is related to the matrix element $M^{2\nu}_{cl}$ governing
the allowed two-neutrino $\beta\beta$ decay. That relation is revealed when these matrix elements are
expressed as functions of the relative distance $r$ between the pair of neutrons that are
transformed into a pair of protons in the $\beta\beta$ decay. Analyzing this relation allows us
to understand the contrasting behavior of these matrix elements when $A$ and $Z$ is changed;
while $M^{0\nu}_{GT}$ changes slowly and smoothly, $M^{2\nu}$ has pronounced
shell effects. We also discuss the possibility of phenomenological determination of the
 $M^{2\nu}_{cl}$ and from them of the $M^{0\nu}_{GT}$ 
 values from the experimental study of the $\beta^{\pm}$ strength functions.
\end{abstract}

\maketitle

\section{Introduction}\label{S:intro} 

Observing  $0\nu\beta\beta$ decay would tell
us that the total lepton number is not a conserved quantity, and that neutrinos
are massive Majorana fermions.  
Answering these questions is obviously a crucial part of the search for the
``Physics Beyond the Standard Model''. Consequently, experimental searches for the 
$0\nu\beta\beta$
decay, of ever increasing sensitivity, are pursued worldwide (for a recent
review of the field, see e.g. \cite{AEE}). However, interpreting existing results 
as a determination
of the neutrino effective mass, and planning new experiments,
is impossible without the knowledge of the corresponding nuclear matrix elements.  
Their determination, and a realistic estimate of their
uncertainty, are therefore an integral part of the problem.

The nuclear matrix elements $M^{0\nu}$ of the  $0\nu\beta\beta$ decay must be evaluated using
tools of nuclear structure theory. Unfortunately, there are no observables that 
could be simply and directly linked to the magnitude of  $0\nu\beta\beta$ nuclear matrix elements
and that could be used to determine them in an essentially model independent way.
In the past, knowledge of the $2\nu\beta\beta$-decay rate,
and therefore of the corresponding matrix elements $M^{2\nu}$, and of the ordinary 
$\beta$ decay $ft$ values and the corresponding beta strength distributions, 
were used to constrain the nuclear model parameters,
in particular when the Quasiparticle Random Phase Approximation (QRPA) was
employed  \cite{us1,us2,us3,SC}. In the present paper we discuss a novel relation between
these nuclear matrix elements. 

Very early, Primakoff and Rosen \cite{Pri59} speculated that since the operators 
governing $0\nu\beta\beta$ and $2\nu\beta\beta$ decays differ by a relatively
gentle radial dependence, approximately of the form $1/r_{ij}$, 
the corresponding matrix elements might be
proportional to each other with the proportionality constant $\sim 1/R$, where $R = 1.2 A^{1/3}$fm
is the nuclear radius. At that time the authors also believed that the  $2\nu\beta\beta$
decay can be treated in closure, thus that the corresponding matrix element is dimensionless,
while in fact the realistic  $2\nu\beta\beta$ matrix element has dimension energy$^{-1}$. Also,
the  $0\nu\beta\beta$ matrix elements are now, by convention, made dimensionless
by including the nuclear radius $R$ as a multiplicative factor, which is compensated
by the factor $R^{-2}$ in the corresponding phase space function $G^{0\nu}$.
 
Modern nuclear structure evaluations of these matrix elements do not support the 
conjecture of proportionality between $M^{0\nu}$ and $M^{2\nu}$. 
The rate of the $2\nu\beta\beta$ decay has been determined experimentally
in many nuclei, and hence the $2\nu\beta\beta$ matrix elements $M^{2\nu}$ are known.
They exhibit pronounced shell effects and vary rather abruptly between
nuclei with different $Z$ and $A$. At the same time, the calculated  $0\nu\beta\beta$ 
nuclear matrix elements, whether based on the QRPA \cite{us1,us2,us3,SC}, nuclear shell model
\cite{Men08,Cau08,Cau08a}, or  the Interacting Boson Model \cite{Bar09}, do not show
such a variability; instead they vary relatively smoothly between nuclei with different $Z,A$.
The reason for the difference is, presumably, 
the very different momentum transfer $q$ involved in these
matrix elements, even though they involve the same initial and final nuclear states.
In the  $2\nu\beta\beta$ decay the momentum transfer $q$ is restricted to $q < Q$, where
$Q$ is the energy difference of the initial and final atomic masses. Hence, the allowed
approximation is valid, $qR \ll 1$, and only the Gamow-Teller operator $\vec{\sigma} \tau_+$
and only the $1^+$ virtual intermediate states, contribute. On the other hand, in the  
$0\nu\beta\beta$ decay the momentum transfer is of the order of the nucleon
Fermi momentum $q \sim 200$ MeV,
$qR \ge 1$, and all $J^{\pi}$ virtual intermediate states can contribute significantly.
Our discussion here sheds more light on the different behavior of the $0\nu\beta\beta$ and
$2\nu\beta\beta$ matrix elements.

It is worthwhile to remember another type of relation, explored in the classic paper by Pontecorvo 
\cite{Pon68}. At that time the available information on $\beta\beta$ decay was based on the 
geochemical determination of the total decay rate, $1/\tau_{tot} = 1/\tau_{0\nu} + 1/\tau_{2\nu}$. 
Since these two modes scale very differently with $Q$ ($\sim Q^5$ for $0\nu$ 
and $\sim Q^{11}$ for $2\nu$)
Pontecorvo suggested that comparing the total lifetimes of two isotopes, $^{130}$Te and $^{128}$Te,
which have very different $Q$ values,
might reveal the presence of the lepton number violating $0\nu$ decay, provided the nuclear matrix
elements of these two isotopes are identical. While the matrix elements of these 
two isotopes  are indeed rather close, they
are not quite the same. Moreover, we know today that the $0\nu$ decay rate is very much smaller, if it
is indeed nonvanishing, than the $2\nu$ decay rate. 

The present paper is structured as follows. In the next section II we describe the formalism
that leads to the relation between the Gamow-Teller part of the $0\nu\beta\beta$ matrix element
and the $2\nu\beta\beta$ matrix element evaluated in the closure approximation. We also discuss
the validity of the closure approximation in the $0\nu\beta\beta$ case. In the following section III
we discuss this novel relation in more detail and show numerous examples. 
In section IV we briefly discuss the issue of quenching of the axial current matrix elements.
While closure is a rather
poor approximation in the $2\nu\beta\beta$ case, we argue
in section V that combining the known lifetimes
with the often measured distribution of the $\beta^-$ and $\beta^+$ strengths constrains the 
$M^{2\nu}_{cl}$ values substantially. We believe that the relation found here allows one to better understand
the different behavior of these matrix elements. We conclude in the last section.

\section{Formalism}\label{S:form}

Assuming that the  $0\nu\beta\beta$ decay is caused by the exchange of the light Majorana neutrinos, the
halflife and the nuclear matrix element are related through
\begin{equation}
\frac{1}{T_{1/2}} = G^{0\nu}(Q,Z) | M^{0\nu} |^2 ~|\langle m_{\beta\beta} \rangle |^2 ~,
\end{equation}
where $ G^{0\nu}(Q,Z)$ is the easily calculable phase space factor, $\langle m_{\beta\beta} \rangle$ is
the effective neutrino Majorana mass whose determination is the ultimate goal of the experiments,
and $ M^{0\nu}$ is the nuclear 
matrix element consisting of Gamow-Teller, Fermi and Tensor parts,
\begin{equation}
 M^{0\nu} = M^{0\nu}_{GT} - \frac{M^{0\nu}_F}{g_A^2} +  M^{0\nu}_T \equiv   
M^{0\nu}_{GT} ( 1 + \chi_F + \chi_T ) ~,
\end{equation}
where $ \chi_F$ and $ \chi_T$ are the matrix element ratios that are smaller than unity and, presumably,
less dependent on the details of the applied nuclear model.
In the following we concetrate on the GT part, $ M^{0\nu}_{GT} $, which can be somewhat symbolically written as
\begin{equation}
 M^{0\nu}_{GT} = \langle f | \Sigma_{lk} \vec{\sigma}_l \cdot  \vec{\sigma}_k \tau_l^+ \tau_k^+ 
H(r_{lk},\bar{E}) | i \rangle ~,
\end{equation}
where $ H(r_{lk},\bar{E})$ is the neutrino potential described in detail below and $r_{lk}$ is the relative
distance between the two neutrons that are trasformed in the decay into the two protons.

\begin{figure}[htb]
\includegraphics[width=.57\textwidth,angle=0]{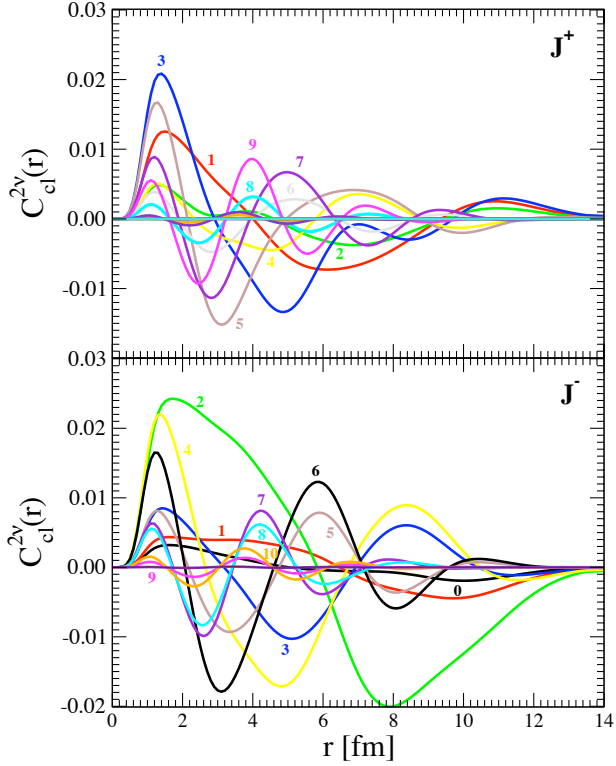}
\caption{Multipole decomposition of $C^{2\nu}_{cl}(r)$ as function of
relative distance of two $\beta$-decaying neutrons in the $^{76}Ge$ nucleus.
Calculation performed for $^{76}$Ge with 23 single particle levels model space.
Positive parity multipoles are shown in the upper panel and the negative parity
ones in the lower panel. (color online)}
\label{fig:C2nu_mult}
\end{figure}

\begin{figure}[htb]
\includegraphics[width=.57\textwidth,angle=0]{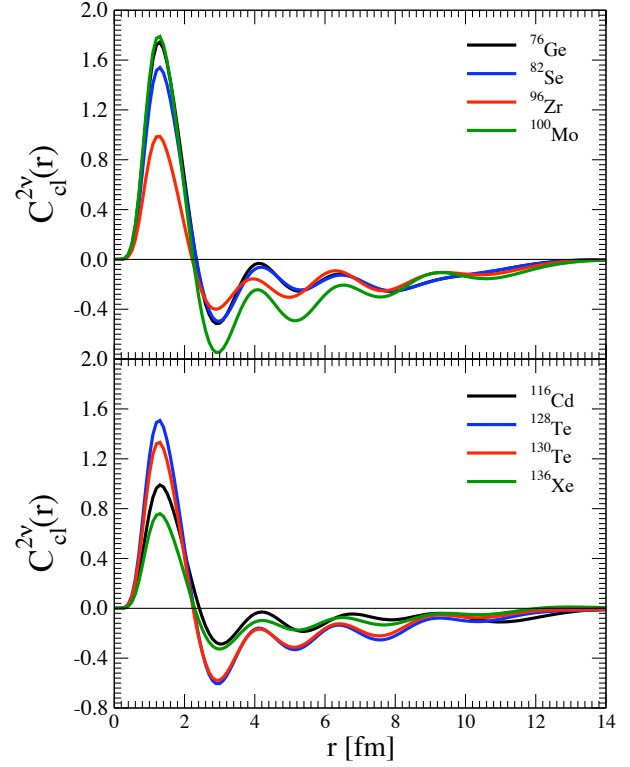}
\caption{$C^{2\nu}_{cl} (r)$ as a function of the relative distance of the decaying neutron pair
for different nuclei. 
(color online)}
\label{fig:C2nu_nucl}
\end{figure}

In Ref. \cite{us3}, based on the QRPA, as well as in Ref. \cite{Men08} based on the nuclear shell model,
the function $C^{0\nu}(r)$ that describes the dependence of the $M^{0\nu}$ on the distance $r_{lk}$
was introduced. Formally, this function can be defined as \cite{EV04}
\begin{equation}
C^{0\nu}_{GT}(r) =  \langle f | \Sigma_{lk} \vec{\sigma}_l \cdot  \vec{\sigma}_k \tau_l^+ \tau_k^+ 
\delta(r - r_{lk}) H(r_{lk},\bar{E}) | i \rangle ~,
\label{eq:C(r)}
\end{equation}
where $\delta(x)$ is the Dirac delta function. Obviously, this function is normalized by
\begin{equation}
 M^{0\nu}_{GT} = \int_0^{\infty} C^{0\nu}_{GT} (r) dr ~,
 \label{eq:C(r)int}
\end{equation}
and has the dimension lenght$^{-1}$. The shape of $C^{0\nu}_{GT}(r)$ 
is very similar in both QRPA and NSM
and in all cases consists of a peak with maximum at $r \sim 1$ fm ending near $r \sim 2.5$ fm,
and of very little contributions for larger values of $r$.

Now lets turn to the case of the $2\nu$ decay mode.
The matrix element $M^{2\nu}$ governing the $2\nu\beta\beta$ decay mode is of the form
\begin{equation}
M^{2\nu} = \Sigma_m \frac{ \langle f || \sigma \tau^+ || m \rangle \langle m || \sigma \tau^+ || i \rangle}
{E_m - (M_i + M_f)/2} ~,
\label{eq:2nu}
\end{equation}
where the sumation extends over all $1^+$ virtual intermediate states. We can introduce also the closure
analog of $M^{2\nu}$, denoted by $M^{2\nu}_{cl}$, by replacing the energies 
$E_m$ by a properly defined average value $\bar{E}_{2\nu}$.
Thus,
\begin{eqnarray}
M^{2\nu}_{cl} \equiv  \langle f | \Sigma_{lk} \vec{\sigma}_l \cdot  \vec{\sigma}_k 
\tau_l^+ \tau_k^+ | i \rangle~,
\nonumber \\
M^{2\nu}_{cl} = M^{2\nu} \times (\bar{E}_{2\nu} - (M_i + M_f)/2) ~.
\label{eq:2nucl}
\end{eqnarray}
In analogy with Eq. (\ref{eq:C(r)}) we can define the new function
\begin{eqnarray}
C^{2\nu}_{cl}(r) =  \langle f | \Sigma_{lk} \vec{\sigma}_l \cdot  \vec{\sigma}_k 
\delta(r - r_{lk}) \tau_l^+ \tau_k^+ | i \rangle ~, \nonumber \\
M^{2\nu}_{cl} = \int_0^{\infty} C^{2\nu}_{cl}(r) dr ~.
\label{eq:2nucl}
\end{eqnarray}
While the matrix elements $M^{2\nu}$ and $M^{2\nu}_{cl}$ get contributions only from the $1^+$
intermediate states, the function $C^{2\nu}_{cl}$ gets contributions
from all intermediate multipoles. 
This is the consequence of the $\delta$ function in the definition of $C^{2\nu}_{cl}(r)$.
When expanded, all multipoles contribute.
Naturally, when integrated over $r$ only the contributions
from the $1^+$ are nonvanishing.
An example of the multipole decomposition of  $C^{2\nu}_{cl}(r)$ is shown in Fig. \ref{fig:C2nu_mult},
and in Fig. \ref{fig:C2nu_nucl} we show the functions $C^{2\nu}_{cl}(r)$ for a variety of $\beta\beta$
decaying nuclei.

For completeness we show here the QRPA formula used for the evaluation of the function
$C^{2\nu}_{cl}(r)$ and its multipole decomposition depicted in Fig.  \ref{fig:C2nu_mult}.
First, the function 
\begin{eqnarray}
& & f^{\mathcal J}_{n,n',p,p'} (r) = \\ \nonumber
& & \langle p(1), p'(2) (r); {\mathcal J} \parallel \vec{\sigma_1} \cdot 
\vec{\sigma_2}  \parallel n(1), n'(2) (r); {\mathcal J} \rangle 
\end{eqnarray}
is introduced where $r$ is the relative distance between the neutrons in the states $n$ and $n'$,
respectively protons in  $p$ and $p'$.  Then, 
the part of $C^{2\nu}_{cl}(r)$ with the multipolarity $J^{\pi}$
is given by
\begin{eqnarray}
& & C^{2\nu}_{cl}(r,J^{\pi})= 
\sum_{k_i,k_f,\mathcal{J}}\sum_{pnp'n'} (-1)^{j_n+j_{p'}+J^{\pi}+{\mathcal J}}   \nonumber \\
& & \times \sqrt{2{\mathcal J}+1} \times\left\{
\begin{array}{c c c}
j_p & j_n & J^{\pi}   \\ 
 j_{n'} & j_{p'} & {\mathcal J}
\end{array}  \right\}  \times f^{\mathcal J}_{n,n',p,p'} (r)  \times  \\
 & & \langle 0_f^+ ||
[ \widetilde{c_{p'}^+ \tilde{c}_{n'}}]_J || J^{\pi} k_f \rangle
\langle  J^{\pi} k_f |  J^{\pi} k_i \rangle
 \langle  J^{\pi} k_i || [c_p^+ \tilde{c}_n]_J || 0_i^+ \rangle .  \nonumber
\end{eqnarray}
Here $k_i$ and $k_f$ are the labels of the excited states with the multipolarity
$J^{\pi}$ in the intermediate nucleus built on the initial and final nuclear
ground states, and $\langle 0_f^+ ||
[ {c_{p'}^+ \tilde{c}_{n'}}]_J || J^{\pi} k_f \rangle$ and  
$\langle  J^{\pi} k_i || [c_p^+ \tilde{c}_n]_J || 0_i^+ \rangle$
are the corresponding QRPA amplitudes.

It is now clear that, by construction,
\begin{equation}
C^{0\nu}_{GT}(r) =  H(r,\bar{E}) \times C^{2\nu}_{cl}(r) ~,
\label{eq:basic}
\end{equation}
which is valid for any shape of the neutrino potential 
$ H(r,\bar{E})$. Thus, if $C^{2\nu}_{cl}(r)$ is known,
$C^{0\nu}_{GT}(r)$ and therefore also $M^{0\nu}_{GT}$ can be easily determined.
The equation (\ref{eq:basic}) represents the basic  
relation between the $0\nu$ and $2\nu$ $\beta\beta$-decay modes that we will explore
further. 

Note that while the function $C^{2\nu}_{cl}(r)$ has a substantial negative tail past $r \sim 2-3$ fm,
these distances contribute very little to $C^{0\nu}_{GT}(r)$. This is a consequence of the
shape of the neutrino potential $H(r,\bar{E})$ that decreases fast
with increasing values of the distance $r$. 
   
\subsection{Neutrino potential}

The neutrino potential $H_{GT}(r,\bar{E})$ governing the Gamow-Teller part of the matrix element
$M^{0\nu}$ is defined as
\begin{eqnarray}
 H_{GT}(r,\overline{E}_{0\nu}) =   \nonumber \\
\frac{2R}{\pi} \int_0^{\infty} j_0(q r)\frac{q}{q + \overline{E}_{0\nu}}
f^2_{FNS}(q^2)g_{HOT}(q^2)dq ~,
\label{eq:pot}
\end{eqnarray}
where 
\begin{equation}
f_{FNS} = \frac{1}{\left(1 + \frac{q^2}{M_A^2}\right)^2}
\end{equation}
takes into account the finite size of the nucleon
and is usually approximated using the above
dipole type form factor with $M_A = 1.09$ GeV \cite{TH95}(varying $M_A$ between 1.0-1.2 GeV makes
little difference). The function $g_{HOT}(q^2)$ includes the 
terms from higher order hadron
currents, namely induced pseudoscalar and weak-magnetism \cite{Sim99}. 
The short range correlations are included using the method of Ref. \cite{Sim09}. The Jastrow-like
two-body function derived there is applied when the radial integrals
in both functions  $C^{0\nu}$ and  $C^{2\nu}_{cl}$ are evaluated; they do not
appear explicitly in eq. (\ref{eq:pot}).

\begin{figure}[htb]
\includegraphics[width=.45\textwidth,angle=0]{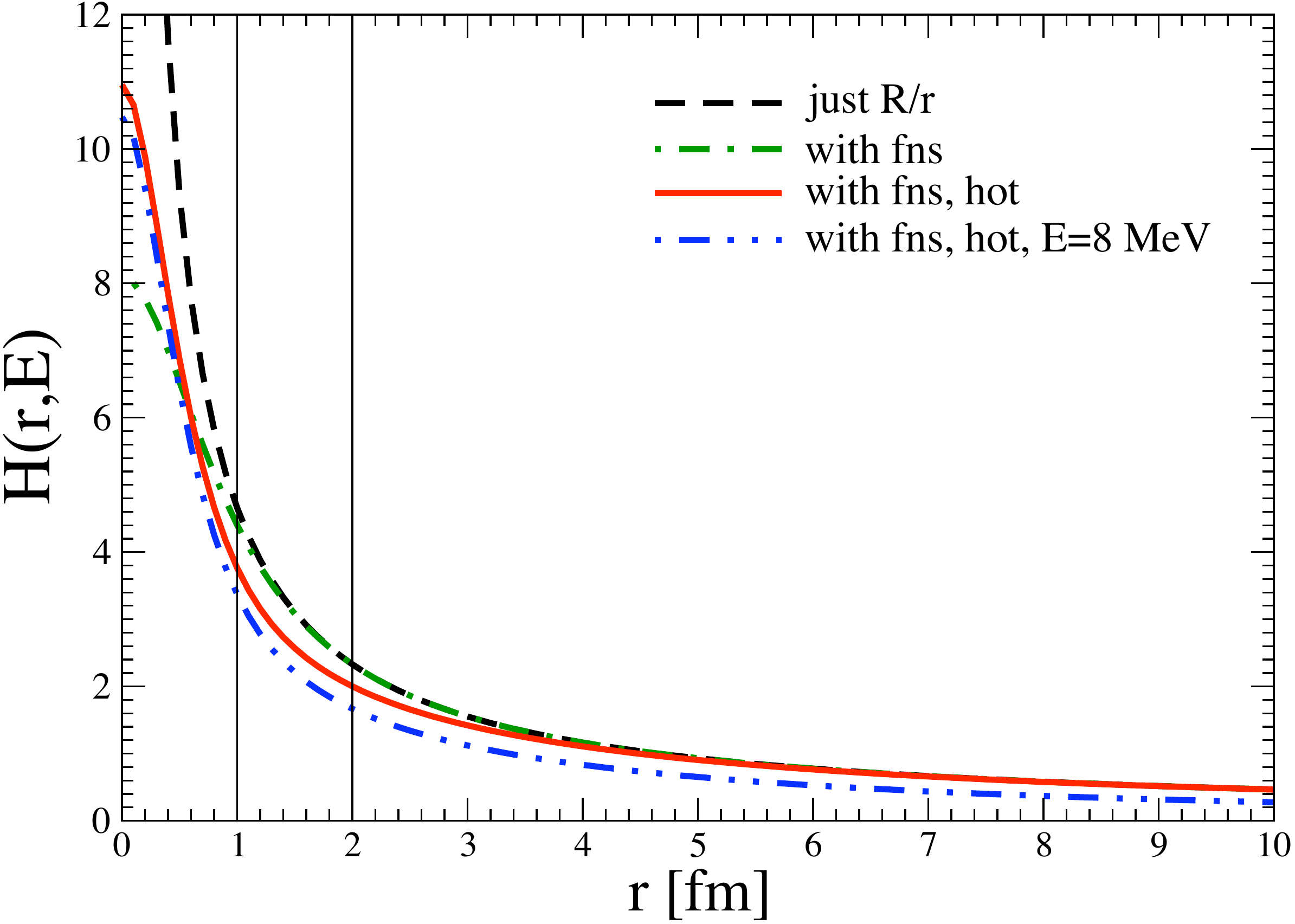}
\caption{The potential $H_{GT} (r, \bar{E})$. Different approximate forms, as well as the exact one, 
are shown (color online).}
\label{fig:pot}
\end{figure}

We show in Fig. \ref{fig:pot} the shape of the potential. When the finite nucleon size,
higher order terms are neglected, and $\bar{E}_{0\nu} = 0$ is assumed, the potential has
Coulomb-like shape $R/r$. The full potential, Eq. (\ref{eq:pot}), however, is finite at $r=0$,
$H(r \rightarrow 0, \bar{E}_{0\nu}=0) = 5M_A R/16$.  Including the higher order currents
and finite $\bar{E}$ in Eq. (\ref{eq:pot}) increases the value of  $H(r=0)$ by $\sim$30\%.

\begin{figure}[htb]
\includegraphics[width=.48\textwidth,angle=0]{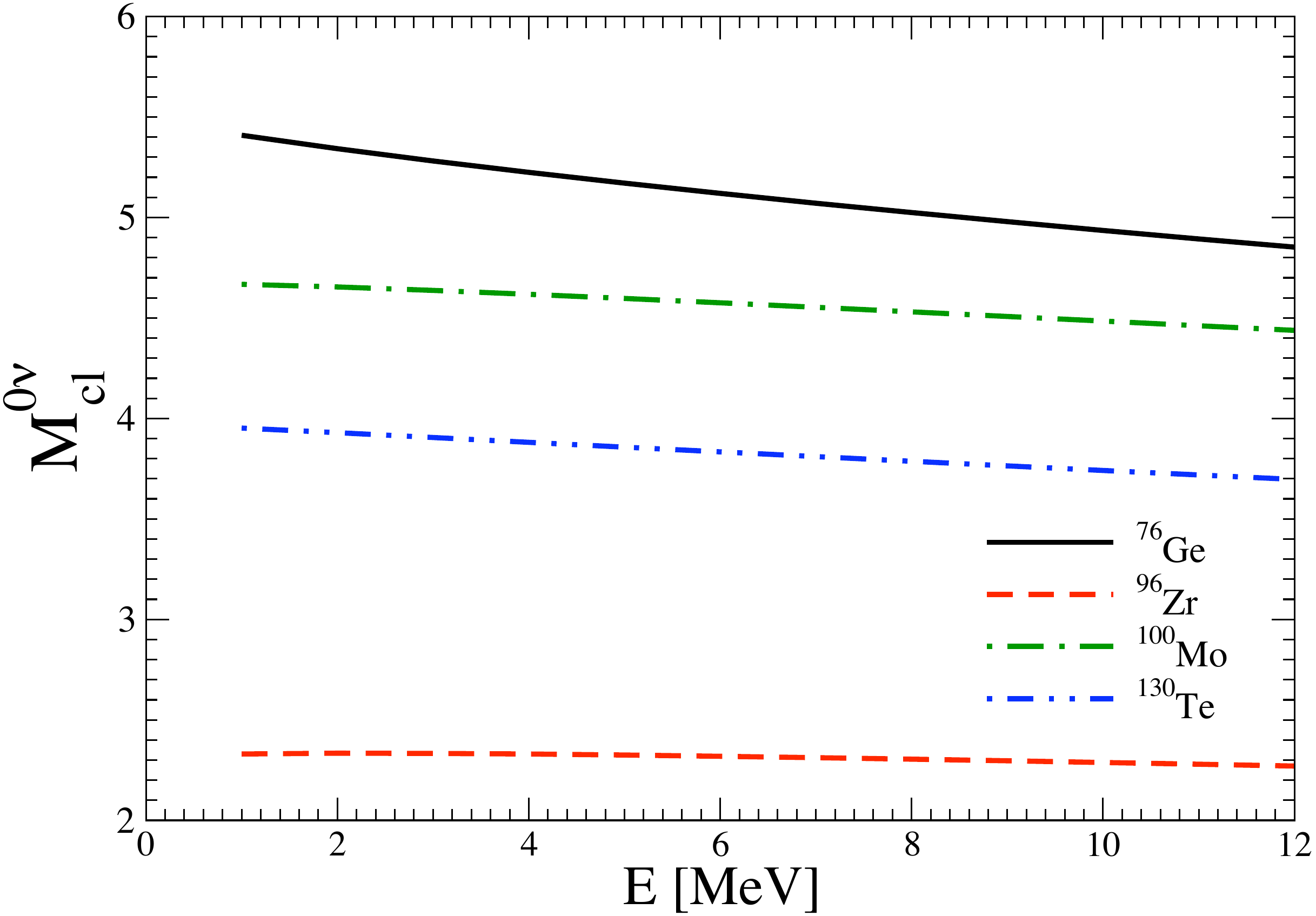}
\caption{
Matrix elements $M^{0\nu}$ for the indicated nuclei evaluated 
in the closure approximation as a function of the assumed average excitation 
energy (color online).
The values of $M^{0\nu}$ obtained without the closure
approximation are 5.24($^{76}$Ge), 2.62 ($^{96}$Zr), 4.99 ($^{100}$Mo), and 4.07
($^{130}$Te).}
\label{fig:cl}
\end{figure}

\subsection{Validity of the closure approximation for the $0\nu\beta\beta$ matrix element}

The closure approximation, i.e. the replacement of the summation over the virtual intermediate states
by  matrix element of a two-body operator, is typically used in the evaluation of the $M^{0\nu}$.
It is worthwhile to test the validity of this approximation. Such test can be conveniently
performed within the QRPA, where the sum over the intermediate states can be easily carried out.
In fact, the calculations performed in Refs. \cite{us1,us2,us3} do not use closure.
In this context one can ask two questions: How good is the closure approximation? And what is
the value of the corresponding average energy?  In Fig. \ref{fig:cl} we illustrate the answers to these
questions. The exact QRPA matrix elements shown in the caption can be compared with the curves obtained by replacing all intermediate energies with a constant $\bar{E}$, 
which is varied there between 0 and 12 MeV.
One can see, first of all, that the $M^{0\nu}$ changes modestly, by less than 10\% when $\bar{E}$
is varied and, at the same time, that the exact results are quite close, but somewhat larger,
than the closure ones. Thus, using
the closure approximation is appropriate for the evaluation of $M^{0\nu}$
even though it slightly underestimates the $M^{0\nu}$ values. However, the corresponding 
uncertainty is not more than the other uncertainties involved.
We compare   the QRPA exact and closure $M^{0\nu}$ for all nuclei of interest in the next section.

\section{Results and Discussion}

We evaluated  the  nuclear matrix elements (NME) $M^{0\nu}_{GT}$ and
$M^{0\nu}$ with and without closure approximation 
using the Quasiparticle Random Phase
Approximation (QRPA). 

For all nuclear systems the  single-particle model space
consisted of $0-5\hbar\omega$ oscillator shells  plus $0i_{11/2}$
and $0i_{13/2}$ levels both for protons and  neutrons
(23 single particle states).
The single particle energies were obtained from
the  Coulomb--corrected Woods--Saxon potential. Two-body interaction G-matrix
elements were derived from the Argonne V18
one-boson exchange potential within the Brueckner theory.
The pairing interaction was
adjusted to fit the empirical pairing gaps \cite{cheo93}. The
particle-particle and particle-hole channels of the G-matrix
interaction of the nuclear Hamiltonian $H$ were renormalized by
introducing the parameters $g_{pp}$ and $g_{ph}$, respectively.
While $g_{ph} = 1.0$ was used throughout,
the particle-particle strength parameter $g_{pp}$
was fixed by the data on the two-neutrino double
beta decay rates \cite{us1,us2,us3} for each nucleus separately.
In the calculation of the $0\nu\beta\beta$-decay NMEs
the two-nucleon short-range correlations
derived from same potential as residual interactions,
namely from the Argonne potential \cite{Sim09}, were applied.
The unquenched value of the axial current coupling constant,
$g_A = 1.269$ was used here. The modifications caused by the
quenching of the weak axial current are discussed in the following
two sections.

On the other hand, the absolute values of $M^{2\nu}_{exp}$ were  
deduced from the averaged values
of $2\nu\beta\beta$-decay half-lives of Ref. \cite{barab}.

In Table \ref{tab:1} we show both the calculated $0\nu\beta\beta$
NMEs evaluated with and without the closure approximation, as well as only
the GT parts of their values.  Also shown are
the experimental $2\nu\beta\beta$-decay NMEs. 
Using the QRPA method the closure matrix elements $M^{2\nu}_{cl} $
were also evaluated. One can 
see that the spread among the candidate nuclei
of the $2\nu\beta\beta$ NMEs is significatly larger when compared
with the spread of the calculated $0\nu\beta\beta$-decay NMEs.
The table also demonstrates that using the closure approximation
for evaluation of $M^{0\nu}$ makes relatively little difference 
 and  that the GT part of $M^{0\nu}$ is dominant
in all considered nuclei.

\begin{table*}[htb]
  \begin{center}
    \caption{The $0\nu\beta\beta$-decay nuclear matrix elements
$|M^{0\nu}_{GT}|$ and $|M^{0\nu}|$ calculated within the QRPA.
For the parameters used, see the text.
The $2\nu\beta\beta$-decay nuclear matrix element $|M^{2\nu}_{exp}|$
were deduced from the avaraged values of the $2\nu\beta\beta$-decay
half-lifes \protect\cite{barab} and $M^{2\nu}_{cl}$ were obtained within the
QRPA. For $^{136}$Xe, where only the upper limit
of the $2\nu$ half-life exists, the range shown covers the range of half-lives
from the experimental limit to infinity. All entries are evaluated with $g_A = 1.269$.}
\label{tab:1}
\renewcommand{\arraystretch}{1.2}
\begin{tabular}{lccccccccc}\hline\hline
NME &  ${^{76}Ge}$ & ${^{82}Se}$ & ${^{96}Zr}$  & ${^{100}Mo}$
    &  ${^{116}Cd}$ & ${^{128}Te}$ & ${^{130}Te}$  & ${^{136}Xe}$ \\ \hline
  & \multicolumn{8}{c}{$2\nu\beta\beta$-decay NMEs}  \\
$|M^{2\nu}_{exp}|~[MeV^{-1}]$ &
0.136 & 0.095 & 0.090 & 0.231 & 0.126 & 0.046 & 0.033 & $(0, 0.031)$ \\
$M^{2\nu}_{cl}$ & 0.099 & -0.126 & -0.802 & -0.933 & 0.059 & -0.462 & -0.464 & (-0.41 -0.25) \\
  & \multicolumn{8}{c}{$0\nu\beta\beta$-decay NMEs within closure approximation}  \\
$|M^{0\nu}_{GT-cl}|$ & 4.12 & 3.61 & 1.89 & 3.72 & 2.77 & 3.63 & 3.09 & $(1.61, 1.83)$ \\
$|M^{0\nu}_{cl}|$   & 5.02 & 4.44 & 2.34 & 4.59 & 3.36 & 4.44 & 3.79 & $(2.00, 2.24)$ \\
  & \multicolumn{8}{c}{$0\nu\beta\beta$-decay NMEs without closure approximation}  \\
$|M^{0\nu}_{GT}|$   & 4.33 & 3.82 & 2.16 & 4.10 & 2.91 & 3.92 & 3.36 & $(1.76, 1.96)$ \\
$|M^{0\nu}|$       & 5.24 & 4.65 & 2.61 & 4.99 & 3.51 & 4.75 & 4.07 & $(2.15,2.38)$ \\
 \hline\hline
\end{tabular}\\
  \end{center}
\end{table*}

The values of $M^{0\nu}$ in Table \ref{tab:1} might be compared with the
corresponding entries in Table II of Ref. \cite{Sim09}. There are small differences
between them caused by several changes made in the present work. We use now
the updated values of $T^{2\nu}_{1/2}$ of Ref. \cite{barab} and the more realistic
$g_A$ = 1.269 instead of 1.25. In evaluating $M^{2\nu}$ we adjust here the energy
denominators such that the first $1^+$ state has the 
experimentally known energy value. Moreover,
the present results are based on the level scheme with just 23 single particle states,
while Ref. \cite{Sim09} uses an average of several sets of single particle energies. 

\begin{figure}[htb]
\includegraphics[width=.47\textwidth,angle=0]{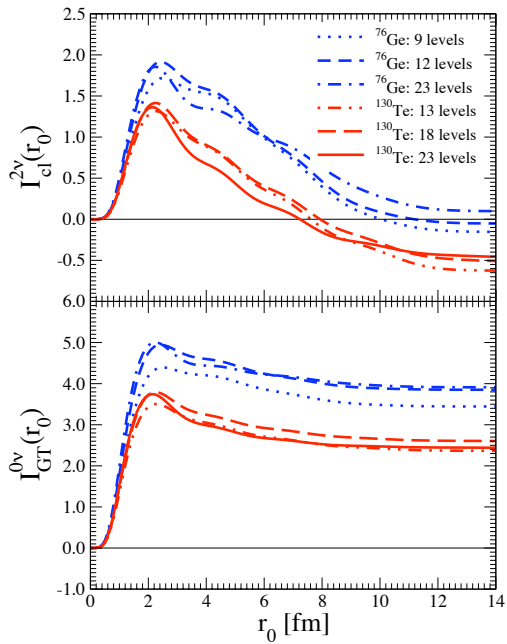}
\caption{Integrals $I^{2\nu}(r_0)$ and $I^{0\nu}_{GT}(r_0)$, Eq.(\ref{eq:ms}),
as function of the upper limit $r_0$. Three different spaces of single-particle 
states are considered, small, medium and large (color online).}
\label{fig:ms}
\end{figure}

Another characteristic feature of the relation
between the $M^{0\nu}_{GT}$ and $M^{2\nu}_{cl}$ is illustrated in Fig. \ref{fig:ms}.
There we show the integrals, i.e. the functions of the upper limit of the integration,
\begin{eqnarray}
I^{2\nu}(r_0) = \int_0^{r_0} C^{2\nu}_{cl} (r) dr~, \nonumber \\
I^{0\nu}_{GT}(r_0) =  \int_0^{r_0} C^{0\nu}_{GT-cl} (r) dr~ .
\label{eq:ms}
\end{eqnarray}
Obviously, ${\rm lim}_{r_0 \rightarrow \infty} I^{2\nu}(r_0) = M^{2\nu}_{cl}$ and
 ${\rm lim}_{r_0 \rightarrow \infty} I^{0\nu}_{GT}(r_0) = M^{0\nu}_{GT}$.

As one can see the integrals $I^{0\nu}_{GT}(r_0)$ saturate for $r_0 \ge 2-3$ fm since the
function $C^{0\nu}_{GT}(r)$ is very small past these values of $r$. On the other
hand, the functions $I^{2\nu}(r_0)$ change drastically, even becoming sometimes 
negative, for $r_0 \ge 2-3$ fm. That is a reflection of the behavior of the function
$C^{2\nu}_{cl}(r)$ that has a substantial tail for $r_0 \ge 2-3$ fm. 
In addition, Fig. \ref{fig:ms} also demonstrates that the corresponding integrals are
almost independent on the number of included single-particle states, as long as
at least two full oscillator shells are taken into account.

Remembering that in a nucleus the average distance between nucleons is $\sim$1.2 fm
we can somewhat schematically separate the range of the variable $r$ in the functions
$C^{0\nu}_{GT}(r)$ and $C^{2\nu}_{cl}(r)$  into the region $r \le$ 2-3 fm governed
by the nucleon-nucleon correlations, while the region $r \ge$ 2-3 fm is governed by
nuclear many-body physics. The integrals in Fig. \ref{fig:ms} demonstrate that 
the matrix elements $M^{0\nu}_{GT}$ are almost independent of the ``nuclear"
region of $r$ and hence one does not expect rapid variations of their value when $A$
or $Z$ of the nucleus is changed. On the other hand, the $2\nu$ closure matrix elements
depend sensitively on that region of $r$ and hence one expects sizable shell effects,
i.e. a significant variation of $M^{2\nu}$ and $M^{2\nu}_{cl}$ with $A$ and $Z$,
in agreement with observations.

\section{Quenching of the axial current matrix elements}

It is well known that Gamow-Teller $\beta$-decay transitions to individual final states 
are noticeably weaker than the theory predicts. That phenomenon is known
as the axial current matrix elements {\it quenching}. The $\beta$-strength functions can 
be studied also with the charge exchange nuclear reactions and a similar effect is 
observed as well. Thus, in order to describe the matrix elements of the operator $\sigma \tau$,
the empirical rule $(\sigma \tau)^2_{eff} \simeq 0.6 (\sigma \tau)^2_{model}$ is usually 
used (see \cite{Ost92,Bro85,Cau94}). Since these operators accompany weak axial current,
it is convenient to account for such quenching by using an effective coupling constant
$g^{eff}_A \sim 1.0$ instead of the true value $g_A = 1.269$.

The evidence for quenching is restricted so far to the Gamow-Teller operator $\sigma \tau$
and relatively low-lying final states.
It is not known whether the other multipole operators associated with the weak axial current
should be quenched as well. In fact, the analysis of the muon capture rates in 
Refs.\cite{Kol00,Zin06}
suggests that quenching is not needed for this process with momentum transfer $q \sim 100$ MeV. 

Since the $2\nu\beta\beta$ decay involves only the GT operators and relatively low-lying
intermediate states, one could expect that the quenching might be involved in that case. Whether
it should be included also for the $0\nu$ mode remains an open question. In the previous
paper, Ref. \cite{us2}, it was shown that by making the adjustment of the particle-particle
coupling strength $g_{pp}$ so that the experimental $2\nu$ halflives are correctly
reproduced, the  predicted $0\nu$ decay rates are affected by the possible 
quenching 
less than the ratio $[(g^{eff}_A/g_A]^4$ might suggest.

Following Ref. \cite{us2} we define the ``quenched" nuclear matrix elements
\begin{equation}
M'^{0\nu} = \left(\frac{g^{eff}_A}{1.269}\right)^2 M^{0\nu}(g^{eff}_A)
\end{equation}
and use the analogous definitions for $M'^{2\nu}_{cl}$, $M'^{0\nu}_{GT}$ and for the
integral $I'^{2\nu}(r_0)$ and $I'^{0\nu}_{GT}(r_0)$ see eq. (\ref{eq:ms}).

We use this definition since the experimental quantities, the halflives $T^{0\nu}$ and
$T^{2\nu}$, are then simply proportional to $1/M'^2$ without the need to modify
the phase space factors $G^{2\nu}$ or $G^{0\nu}$ when a different value of
$g^{eff}_A$ is used.  Note that as a consequence of our choice of renormalization
of the particle-particle coupling constant $g_{pp}$ the $2\nu\beta\beta$ matrix elements
$M^{2\nu}$ by definition become independent of $g^{eff}_A$ and thus $M^{2\nu} = M'^{2\nu}$.

\begin{figure}[htb]
\includegraphics[width=.47\textwidth,angle=0]{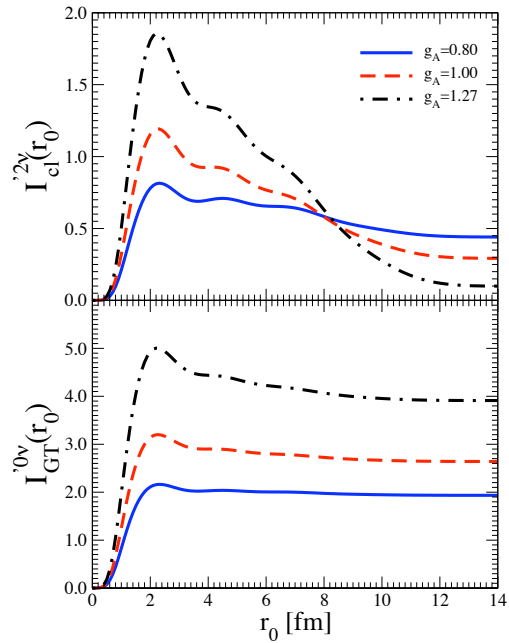}
\caption{ The running sums of $I'^{2\nu}$ (upper panel)
 and $I'^{0\nu}_{GT}$ (lower panel) for $^{76}$Ge and different effective
 values of $g_A$. (color online).}
\label{fig:ga}
\end{figure}

In Fig. \ref{fig:ga} we show the integrals $I'^{2\nu}_{cl}(r_0)$ and $I'^{0\nu}_{GT}(r_0)$
 for the case of the decay of $^{76}$Ge and three values of $(g^{eff}_A)$.
One can see that in the case of $M'^{2\nu}_{cl}$ not only does the final value depend on 
$(g^{eff}_A)$, but it affects the dependence on the distance $r_0$ as well. With the standard
$g_A = 1.27$ the peak at $r_0 \sim 2$ fm is compensated by the long tail 
extending to much larger $r_0$, while for the heavily quenched $g^{eff}_A = 0.8$ the
$I'^{2\nu}(r_0)$ almost saturates at the much smaller values of $r_0$. In contrast,
the integrals $I'^{0\nu}_{GT}(r_0)$ saturate at $r_0 \sim 3$ fm for all considered values of $g^{eff}_A$.

It was shown in Ref.\cite{us2} that with $g^{eff}_A = 1.0$ the full matrix elements $M'^{0\nu}$
are reduced by 10-15\% compared to their value with $g_A = 1.25$ used in that paper.
Here we use the more correct $g_A = 1.269$, for adjustment of the particle-particle coupling
constant $g_{pp}$ we use the $2\nu$ halflives of Ref. \cite{barab} that differ slightly
from the halflives used in \cite{us2}, and the treatment of the short-range correlations
is now based on the Ref. \cite{Sim09} while in \cite{us2} it was based on the
phenomenological Jastrow-type function of Ref. \cite{Mil76}. In the present work
the matrix elements $M'^{0\nu}$ are 20-30\% smaller with $g^{eff}_A = 1.0$ than
with $g_A = 1.269$. Similar effects are also visible in Table II of Ref. \cite{Sim09}.

\section{Determination of the matrix element $M^{2\nu}_{cl}$}

While the nuclear matrix elements $M^{2\nu}$ are simply related to the $2\nu$ half-life 
$T^{2\nu}_{1/2}$, and are therefore known for the nuclei in which $T^{2\nu}_{1/2}$ has
been measured, the closure matrix elements  $M^{2\nu}_{cl}$ need be determined separately.
There are several ways how to accomplish this task:
\begin{enumerate}
\item Rely on a nuclear model (e.g. QRPA or nuclear shell model), adjust parameters
in such a way that the experimental value of $M^{2\nu}$ is correctly reproduced,
and use the model to evaluate $M^{2\nu}_{cl}$. (In QRPA the usual adjustment 
is the renormalization of the particle-particle coupling constant $g_{pp}$ so that
the $T^{2\nu}_{1/2}$ is correctly reproduced.) This procedure is used in Table \ref{tab:1}.
\item Use the measured $\beta^-$ and $\beta^+$ strength functions and assume 
coherence (i.e. same signs) among states with noticeable strengths in both channels.
In this way an upper limit of  $M^{2\nu}_{cl}$ can be obtained.
\item Finally, one could invoke the so called ``Single state dominance hypothesis"
\cite{Abad84} according to which the sum in the eqs. (\ref{eq:2nu}) and (\ref{eq:2nucl})
is exhausted by its first term. The measured $\beta$ decay and $EC$ $ft$ values then
make it possible to determine both  the $M^{2\nu}$ and $M^{2\nu}_{cl}$.
\end{enumerate}
Obviously, none of these methods is exact, but their combination has, perhaps, a chance
of constraining the value of $M^{2\nu}_{cl}$ substantially. Examples of application
of the latter two items are shown in Table \ref{tab:2}. That method can be used,
obviously, only for the nuclei where the corresponding experimental data are available.

\begin{table*}[htb]  
  \begin{center}  
    \caption{ The $2\nu\beta\beta$-decay closure nuclear matrix element
$|M^{2\nu}_{cl}|$ evaluated using the 
Single State Dominance Hypothesis (SSD) and with help of the measured
$\beta^{\pm}$ strengths in charge exchange reactions (ChER). The adopted values of the 
$2\nu\beta\beta$-decay half-times $T^{2\nu-exp}_{1/2}$, taken from Ref.
 \cite{barab}  are also shown.
In the ChER case the matrix elements $|M^{2\nu}|$ and $M^{2\nu}_{cl}$ have been 
determined  by assuming  equal phases 
for its each individual contribution. }  
\label{tab:2}  
\renewcommand{\arraystretch}{1.2}  
\begin{tabular}{lccccccc}\hline\hline
 & &  \multicolumn{3}{c}{$SSD$} &  \multicolumn{3}{c}{$ChER$} \\
\cline{4-5} \cline{7-8}
Nucleus &  $T^{2\nu-exp}_{1/2}$ [y] &  &
$|M^{2\nu}|$ $[MeV^{-1}]$ & $|M^{2\nu}_{cl}|$ &  & 
$|M^{2\nu}|$ $[MeV^{-1}]$ & $|M^{2\nu}_{cl}|$  \\ \hline
$ ^{48}Ca$  & $4.4\times 10^{19}$   &  & -                      & -    & & 0.083 & 0.220 \cite{Yako}     \\
$ ^{76}Ge$  & $1.5\times 10^{21}$ &  & -                        & -    & & 0.159 & 0.522 \cite{Grewe}   \\
$ ^{96}Zr$  & $2.3\times 10^{19}$  &  & -                        & -    & &   -   & 0.222 \cite{Dohmann}  \\
$^{100}Mo$  & $7.1\times 10^{18}$  &  & 0.208 & 0.350 \cite{Domin}  &  &   -                  &  -     \\
$^{116}Cd$  & $2.8\times 10^{19}$  &  & 0.187 & 0.349 \cite{Domin}  &  & 0.064 & 0.305 \cite{Rakers}   \\
$^{128}Te$  & $1.9\times 10^{24}$  &  & 0.019 & 0.0327 \cite{Domin} &  &  -                   &  -     \\
\hline\hline
\end{tabular}\\
  \end{center}  
\end{table*}

 Comparison of the NMEs $M^{2\nu}_{exp}$ and $M^{2\nu}_{cl}$ in Table \ref{tab:1}
tells us, right away that, at least within the QRPA, the summation in the Eqs. (\ref{eq:2nu})
and (\ref{eq:2nucl}) contains both positive and negative parts (see also Fig. \ref{fig:ms}).
This is obviously so since for most nuclei the quantity $\bar{E}_{2\nu} - (M_i + M_f)/2$ in
Eq. (\ref{eq:2nucl}) becomes negative, while each of the denominators in the Eq. (\ref{eq:2nu})
is positive. Hence, we cannot expect good agreement between the $M^{2\nu}_{cl}$ from QRPA
and those from the items 2. and 3. above. 
And, moreover, we cannot expect that SSD is a valid hypothesis for all candidate
nuclei. Comparison of the corresponding entries in
Tables \ref{tab:1} and \ref{tab:2} confirms that expectation.

\begin{figure}[htb]
\includegraphics[width=.55\textwidth,angle=0]{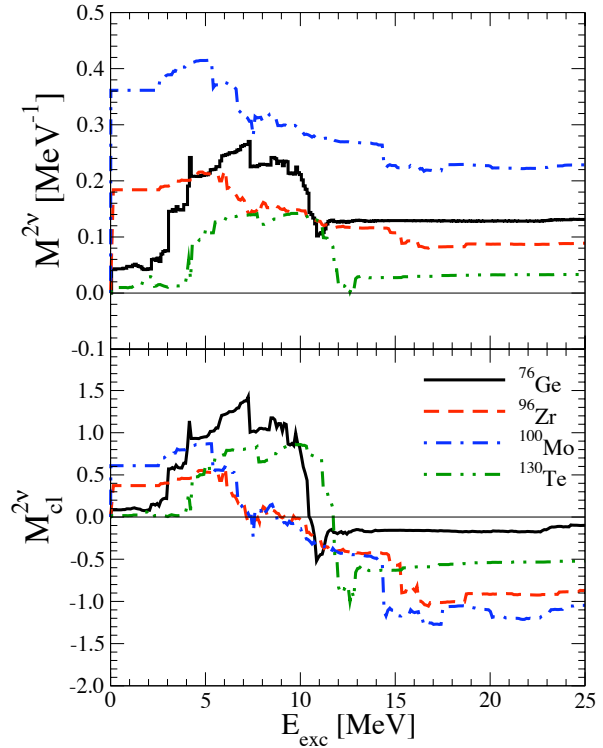}
\caption{ The running sums of $M^{2\nu}$ (upper panel)
 and $M^{2\nu}_{cl}$ (lower panel) for selected nuclei. (color online).
 $g_A = 1.269$ was used.}
\label{fig:stair}
\end{figure}

Since there is a substantial experimental activity devoted to the determination of the
$\beta^{\pm}$ strengths, it is worthwhile to examine in more detail the somewhat unexpected 
finding that in many cases $M^{2\nu}$ and $M^{2\nu}_{cl}$ have opposite signs. Obviously,
this has to do with the different weight of the corresponding terms in the Eqs.  (\ref{eq:2nu})
and (\ref{eq:2nucl}). We plot in Fig. \ref{fig:stair} the corresponding running sums as a function 
of the excitation energy in the intermediate nucleus. 
One can see that the negative values of $M^{2\nu}_{cl}$ arise from excitation energies
$E_{ex} > 10$ MeV that are difficult to explore experimentally.

 The negative contributions to $M^{2\nu}$ and $M^{2\nu}_{cl}$ from higher
 excitation energies cause in several nuclei  even the reversal of the sign of   $M^{2\nu}_{cl}$ to 
 the negative one. While, clearly, there is a substantial $\beta^-$ strength at these excitation
 energies, QRPA predicts that there is a sufficient $\beta^+$ strength there as well, leading
 to the reduction of the $M^{2\nu}$ and $M^{2\nu}_{cl}$ visible in Fig. \ref{fig:stair}.
 Our QRPA calculations suggest that about 0.2 units of the B(GT) $\beta^+$
 strength is distributed among states with $E_{ex} \ge 10$ MeV in all considered
 nuclei.  Such $\beta^+$ strength has not been observed experimentally so far. It remains to be
 seen whether it exists at all, or is hidden in the ``grass", i.e. distributed among many weak
 states that escape identification. Until this dilemma is resolved we cannot decide whether
 the  closure matrix elements $M^{2\nu}_{cl}$ in Table I are realistic or not.
 
 In the previous section we discussed the phenomenon of quenching of the axial current
 matrix elements. Fig. \ref{fig:ga} suggests that using the effective $g^{eff}_A < 1.27$ reduces
 the negative contribution of the higher lying $1^+$ states to the matrix element  
 $M'^{2\nu}_{cl}$. To see how large that effect might be we performed QRPA calculation
 with $g^{eff}_A = 0.9$ based on the empirical evidence that the degree of quenching increases
 with $A$.  The resulting quenched matrix elements $M'^{2\nu}_{cl}$ are shown in 
 Table \ref{tab:3}. While, as remarked earlier, it is unknown whether all mutipoles are
 affected by the axial current quenching, not only the GT $1^+$ states, we 
 nevertheless show in the
 same Table the values of quenched $M'^{0\nu}_{GT}$ and of the full NME $M'^{0\nu}$
 with and without the closure approximation. 
 If quenching would not affect these matrix elements, their magnitude would be enhanced
 by the factor $(1.269/0.9)^2 \sim 2$ making them substantially larger than the values in 
 Table \ref{tab:1}.
 
\begin{table*}[htb]
\begin{center}
 \caption{The experimental values $M'^{2\nu}_{exp}$ and the QRPA values of 
 $M'^{2\nu}_{cl}$ evaluated with quenching at $g_A = 0.9$. The corresponding
 quenched values of the $0\nu\beta\beta$ matrix elements are also shown.}
\label{tab:3}
\renewcommand{\arraystretch}{1.2}
\begin{tabular}{lccccccccc}\hline\hline
NME &  ${^{76}Ge}$ & ${^{82}Se}$ & ${^{96}Zr}$  & ${^{100}Mo}$
    &  ${^{116}Cd}$ & ${^{128}Te}$ & ${^{130}Te}$  & ${^{136}Xe}$ \\ \hline
  & \multicolumn{8}{c}{$2\nu\beta\beta$-decay NMEs}  \\
$|M'^{2\nu}_{exp}|~[MeV^{-1}]$ &
0.136 & 0.095 & 0.090 & 0.231 & 0.126 & 0.046 & 0.033 & $(0, 0.031)$ \\
$M'^{2\nu}_{cl}$ & 0.336 & 0.100 & -0.210 & -0.205 & 0.179 & -0.146 & -0.169 & (-0.25,  -0.047) \\
& \multicolumn{8}{c}{$0\nu\beta\beta$-decay quenched NMEs within closure approximation}  \\
$M'^{0\nu}_{GT-cl}$ & 2.50 & 2.10 & 1.20 & 2.23 & 1.63 & 2.09 & 1.77 & (0.86, 1.08)  \\
$M'^{0\nu}_{cl}$   & 3.82 & 3.25 & 1.91 & 3.49 & 2.45 & 3.28 & 2.82 & (1.43, 1.72) \\
& \multicolumn{8}{c}{$0\nu\beta\beta$-decay quenched NMEs without closure approximation}  \\
$M'^{0\nu}_{GT}$ & 2.59 & 2.20 & 1.33 & 2.45 & 1.71 & 2.23 & 1.91 & (0.93, 1.14)  \\
$M'^{0\nu}$   & 3.90 & 3.34 & 2.05 & 3.71 & 2.53 & 3.44 & 2.96 & (1.51, 1.78) \\
 \hline\hline
\end{tabular}\\
  \end{center}
\end{table*}

Since our goal is the determination of the GT part of
the $0\nu\beta\beta$ matrix element $M^{0\nu}_{GT}$, a priori the knowledge of the
$M^{2\nu}_{cl}$, which depends only on the $1^+$ intermediate states, is insufficient. 
According to the Eq. (\ref{eq:basic}) we need
for that purpose  the function $C^{2\nu}_{cl}(r)$ that depends,
in principle, on all intermediate multipoles. However, is we could use the expansion
of the spherical Bessel function $j_0(qr)$ in Eq.(\ref{eq:pot}) in powers of $qr$ and keep
just the first term, the neutrino potential $H(r,\bar{E})$ would be represented by a constant
and the Eq.(\ref{eq:basic}) would predict a simple proportionality between $M^{2\nu}_{cl}$
and $M^{0\nu}_{GT}$. However, such an expansion does not work. In reality in Eq. (\ref{eq:pot})
$qr \ge 1$ and we cannot approximate the neutrino potential $H(r,\bar{E})$ by its value at $r=0$.
Hence, we do not expect a proportionality between $M^{2\nu}_{cl}$ and $M^{0\nu}_{GT}$ and
the QRPA evaluation supports this conclusion.

\section{Conclusions}

Since the nuclear matrix elements $M^{0\nu}$ must be determined theoretically, it is of
obvious interest to search for any relation between their numerical values and other
quantities that are either known from experiments or at least constrained by them.
Here we describe such a relation between the dominant Gamow-Teller part 
$M^{0\nu}_{GT}$ of $M^{0\nu}$ and the matrix element $M^{2\nu}_{cl}$  of the
observed $2\nu\beta\beta$-decay evaluated in the closure approximation.
The relation is based on the evaluation of the auxiliary functions $C^{0\nu}_{GT}(r)$
and $C^{2\nu}_{cl}(r)$ that describe the dependence of the corresponding
nuclear matrix elements on the distance $r$ between the pair of neutrons that is
transformed in the $\beta\beta$ decay into a pair of protons.
Thus (see Eqs. (\ref{eq:C(r)int}), (\ref{eq:2nucl}) and (\ref{eq:basic}))
\begin{eqnarray}
& &M^{0\nu}_{GT} = \int_0^{\infty} C^{0\nu}_{GT} (r) dr ~, 
~M^{2\nu}_{cl} = \int_0^{\infty} C^{2\nu}_{cl}(r) dr ~, \nonumber \\
& & {\rm and}~C^{0\nu}_{GT}(r) =  H(r,\bar{E}) \times C^{2\nu}_{cl}(r) ~,
\end{eqnarray}
represents the required relation.

However, while the matrix elements $M^{2\nu}$ and $M^{2\nu}_{cl}$ depend only
on the transition strengths and energies of the $1^+$ virtual intermediate states
(they are pure $GT$ quantities), the function $C^{2\nu}_{cl}(r)$ gets contribution
from all multipoles. Thus, the relation that we found is an indirect one; even if
$M^{2\nu}_{cl} $ would be precisely known, the evaluation of the function
$C^{2\nu}_{cl}(r)$ requires additional nuclear theory input.

Nevertheless, the relation in Eq. (\ref{eq:basic}) allows us to obtain a better insight
into the problem of the $A$ and $Z$ dependence of the matrix elements
$M^{2\nu}$ and $M^{0\nu}$. While the known $M^{2\nu}$ have a strong 
shell dependence, the calculated $M^{0\nu}$ vary much less. Analysis
of the functions $C^{2\nu}_{cl}(r)$ and $C^{o\nu}_{GT}(r)$ makes it possible
to better understand where this fundamental difference comes from. 

We show that, so far, the QRPA values of closure approximation $M^{2\nu}_{cl}$ 
matrix elements do not 
agree well with the same quantities based on the measured $\beta^-$ and $\beta^+$
strength functions and on the assumption of coherence (i.e. same sign) of contributions
of individual states. Until this discrepancy is resolved, it is difficult to employ $M^{2\nu}_{cl}$
in order to constrain the magnitude of the $0\nu\beta\beta$ matrix elements $M^{0\nu}_{GT}$.

\section*{Acknowledgments}

 Useful discussions with Kazuo Muto are appreciated. The work of P.V.\ was
partially supported by the US Department of Energy under Contract No.\
DE-FG02-88ER40397. A.F., R.H. and F.\v S acknowledge 
the support in part by the DFG project 436 SLK 17/298,  the Transregio Project 
TR27 "Neutrinos and Beyond" and by the VEGA 
Grant agency  under the contract No.~1/0249/03. 


\end{document}